\documentclass[12pt]{article}
\usepackage{epsfig}
\begin{document}
\begin{titlepage}
\begin{center}
{\hbox to\hsize{\hfill UMD-PP-02-036}}

\vspace{4\baselineskip}

\textbf{\Large 
Leptogenesis in models with multi-Higgs bosons }
 
\bigskip
\bigskip
\vspace{2\baselineskip}

\textbf{Takeshi Fukuyama%
\footnote{E-Mail: fukuyama@physics.umd.edu}} \\ 
\bigskip
\textit{\small 
Department of Physics, Ritsumeikan University, Kusatsu, 
Shiga 525-8577, Japan \\
and \\
Department of Physics, University of Maryland, 
College Park, MD 20742, USA 
}
\vspace{2\baselineskip}

\textbf{Nobuchika Okada%
\footnote{E-Mail: okadan@physics.umd.edu} }\\
\bigskip
\textit{\small
Department of Physics, University of Maryland,  
College Park, MD 20742, USA}

\vspace{3\baselineskip}

\textbf{Abstract}\\
\end{center}
\noindent
We study the leptogenesis scenario in models 
 with multi-Higgs doublets. 
It is pointed out that the washing-out process 
 through the effective dimension five interactions, 
 which has not been taken into account seriously 
 in the conventional scenario, 
 can be effective, and the resultant baryon asymmetry 
 can be exponentially suppressed. 
This fact implies new possible scenario 
 where the observed baryon asymmetry is the remnant of 
 the washed out lepton asymmetry which was originally much larger 
 than the one in the conventional scenario. 
Our new scenario is applicable to some neutrino mass matrix models  
 which predict too large CP-violating parameter 
 and makes them viable through the washing-out process.   
\end{titlepage}
%
\setcounter{footnote}{0}
\newpage

The origin of the observed baryon asymmetry,  
 the ratio of the baryon density to the entropy density 
\begin{eqnarray}
Y_B = \frac{n_B-n_{\bar{B}}}{s} = (0.6 - 1) \times 10^{-10} 
\; ,  \nonumber 
\end{eqnarray} 
 is one of the major problems in cosmology. 
Various scenarios of baryogenesis have been discussed 
 \cite{riotto}, and leptogenesis \cite{fukugita} \cite{LGreview}
 is one of the attractive scenarios. 
Providing $U(1)_{B-L}$ breaking in the original theory, 
 lepton number is generated through the CP-violating 
 out-of-equilibrium decay 
 of the right-handed Majorana neutrinos. 
A part of this lepton number is converted 
 into the baryon number 
 via the (B+L) violating sphaleron process 
 being in thermal equilibrium \cite{kuzmin}. 
As a result, the baryon asymmetry 
 in the universe is generated. 

The leptogenesis is also an interesting scenario 
 in the following point of view. 
Amount of the generated lepton asymmetry (baryon asymmetry) 
 is related to the neutrino Dirac mass matrix, 
 and thus related to neutrino oscillation data \cite{data} 
 through the see-saw mechanism \cite{seesaw} 
 which naturally explains the smallness of the neutrino masses. 
For detailed discussion, some concrete models are necessary. 
There have been many works on the models of neutrino mass matrix, 
 where the relationship between the neutrino oscillation data 
 and the observed baryon asymmetry are discussed in detail. 

While the conventional leptogenesis scenario has been discussed 
 for one Higgs doublet model, 
 in this letter, we study the leptogenesis 
 in the models with (up-type) multi-Higgs doublets. 
In multi-Higgs models, 
 one-to-one correspondence between 
 the Yukawa couplings and 
 the neutrino Dirac mass matrix 
 is no longer valid. 
This fact leads to new possible scenario of the leptogenesis, 
 where the washing-out process 
 through the dimension five interactions 
 plays the crucial role for the resultant baryon asymmetry. 
Since the effect of this washing-out process 
 has not been considered seriously, 
 the scenario is worth investigating.  

Let us first briefly review the conventional leptogenesis scenario. 
In the following, our discussion is always 
 based on the effective Lagrangian 
 at energies lower than the right-handed neutrino masses such that 
\begin{eqnarray}
{\cal L}_{N}=-h_{ij} 
\overline{l_{L,i}} \phi N_j\nonumber 
-\frac{1}{2} \sum_{i} 
 \overline{ N^C_i} M_i N_i + h.c. \; , 
\end{eqnarray} 
where $i,j=1,2,3$ denote the generation indeces, 
 $h$ is the Yukawa coupling, 
 $l_L$ and $\phi$ are the lepton and the Higgs doublets, 
 respectively, and 
 $M_i$ is the lepton-number-violating mass term  
 of the right-handed neutrino $N_i$ 
 (we are working on the basis of 
 the right-handed neutrino mass eigenstates). 
For simplicity, we assume 
 the hierarchy among the right-handed neutrino masses, 
 $M_1 \ll M_2 \ll M_3$, in the following. 

The lepton asymmetry in the universe is generated 
 by CP-violating out-of-equilibrium decay 
 of the heavy neutrinos,  
 $N \rightarrow l_L \overline{\phi}$ 
 and $ N \rightarrow \overline{l_L} \phi $. 
The leading contribution 
 is given by the interference between 
 the tree level and one-loop level decay amplitudes, 
 and the CP-violating parameter is found to be \cite{epsilon}%
 \footnote{ 
 Through out this letter, our notations  
 are all based on Ref.~\cite{plumacher}. }
\begin{eqnarray}
\epsilon = 
 \frac{1}{8\pi (h^\dagger h)_{11}}  
 \sum_{j}\mbox{Im} \left[ (h^\dagger h)_{1j}^2   \right]
 \left\{ f(M_j^2/M_1^2)
+ 2 g(M_j^2/M_1^2) \right\} \; .
\label{epsilon}
\end{eqnarray}
Here $f(x)$ and $g(x)$ correspond to 
 the vertex and the wave function corrections, 
\begin{eqnarray}
f(x)&\equiv& \sqrt{x} \left[  
1-(1+x)\mbox{ln} \left(\frac{1+x}{x} \right) \right] \; ,
 \nonumber \\
g(x)&\equiv& \frac{\sqrt{x}}{2(1-x)}   \; ,  
\end{eqnarray}
 respectively, and both are reduced to 
 $\sim -\frac{1}{2 \sqrt{x}}$ for $ x \gg 1$. 
We have assumed that the lightest $N_1$ decay 
 dominantly contributes to the resultant lepton asymmetry. 
In fact, this is confirmed by numerical analysis 
 in the case of hierarchical right-handed neutrino masses 
 \cite{plumacher}. 
Using the above $\epsilon$, 
 the generated $Y_{B}$ is described as
\begin{eqnarray}
Y_{B}  \sim  \frac{\epsilon}{g_*}  d \; , 
\end{eqnarray}
where $g_* \sim 100$ is the effective degrees of freedom 
 in the universe at $T \sim M_1$, 
 and $ d \leq 1 $ is so-called the dilution factor. 
This factor parameterizes how the naively 
 expected value $Y_B \sim \epsilon/g_*$ is reduced 
 by washing-out processes. 

We can classify the washing-out processes 
 into two cases with and without the external leg 
 of the heavy right-handed neutrinos, respectively. 
The former includes the inverse-decay process 
 and the lepton-number-violating scatterings 
 mediated by the Higgs boson \cite{luty} such as 
 $N + \overline{l_L} \leftrightarrow \overline{q_R} + q_L$,  
 where $q_L$ and $q_R$ are quark doublet and singlet, 
 respectively. 
The latter case is the one induced 
 by the effective dimension five interaction, 
\begin{eqnarray} 
 {\cal L}_N \sim \frac{h^2}{M}l l \phi \phi \; , 
\label{4point}
\end{eqnarray}
 after integrating out the heavy right-handed neutrinos. 
Here we have described the interaction symbolically 
 omitting the generation indeces. 
This term is nothing but the one 
 providing the see-saw mechanism \cite{seesaw}. 
The importance of this interaction 
 was discussed in \cite{kolb},  
 where the interaction was shown to be necessary 
 to avoid the false  generation of the lepton asymmetry 
 in thermal equilibrium. 
While numerical calculations  \cite{plumacher} \cite{luty}
 are necessary in order to evaluate the dilution factor precisely, 
 $Y_B \sim \epsilon/g_{*}$ roughly gives a correct answer, 
 and the washing-out process is mostly not so effective. 
Note that this is the consequence 
 from the current neutrino oscillation data 
 as explained in the following. 

The condition for a washing-out process to be effective 
 is roughly given by 
\begin{eqnarray}
 \Gamma(T\sim M_1)  \geq H(T \sim M_1) 
 \sim \sqrt{g_{*}} \frac{M_1^2}{M_P}  \; , 
\end{eqnarray}
 where $\Gamma$ denotes the decay width or 
 the thermal-averaged cross section times number density 
 for the lepton-number-violating scatterings, 
 $H$ is the Hubble parameter, 
 and $M_P \sim 10^{19}$ GeV is the Planck mass. 
This condition leads to the lower bound on 
 the light Majorana neutrino mass eigenvalues.%
\footnote{
See Eqs.~(\ref{PDbound}) and (\ref{PNbound}) 
 for more precise  evaluation. 
}
For simplicity, let us assume 
 $(h^\dagger h)_{ii} v^2 / M_i = (m_D^\dagger m_D)_{ii}/M_i
 \sim m_{\nu i} $ through the see-saw mechanism, 
 where $v =174$ GeV is the vacuum expectation value (VEV) 
 of the Higgs doublet, 
 and $m_{\nu i}$ is the mass eigenvalue of 
 the light Majorana neutrino in the $i$-th generation. 

Considering that the (inverse) decay width is given by 
\begin{eqnarray} 
 \Gamma_D \sim c \; (h^\dagger h)_{11} M_1 
 \sim c \; m_{\nu 1} M_1^2/v^2 \; , 
\end{eqnarray} 
where $c \sim 10^{-2}$ is the phase space factor, 
 we can find $ m_{\nu 1} \geq {\cal O}(10^{-3} \mbox{eV})$.
Although estimation for the Higgs mediated scatterings is 
 much more complicated, we obtain the similar result. 
To evaluate the condition for the washing-out process 
 due to the dimension five interaction 
 needs knowledge of the flavor structure 
 in the neutrino sector. 
For simplicity, assume that the dominant contributions 
 are coming from each single $N_i$ exchange for $i$ fixed. 
In this case, we find the cross section 
\begin{eqnarray} 
 \sigma \sim c \; \sum_i (h^\dagger h)_{ii}^2/M_i^2 
 \sim c \; \sum_i m_{\nu i}^2/v^4 \; , 
\end{eqnarray} 
and the thermal-averaged cross section times number density 
 is given by $\Gamma(T \sim M_1) \sim \sigma M_1^3 $. 
This leads to the bound such as
 $ \sum_i m_{\nu i}^2 \geq {\cal O}(\mbox{eV}^2) 
 \left( \frac{10^{10}\mbox{GeV}}{M_1} \right) $. 
Note that this result is incompatible 
 with the neutrino oscillation data \cite{data} 
 in the case of hierarchical light neutrino masses, 
 since $ \sum_i m_{\nu i}^2 \sim  
 \Delta m_\oplus^2 \sim 10^{-3}~\mbox{eV}^2$, 
 where $\Delta m_\oplus^2$ is the neutrino oscillation parameter 
 relevant for the atmospheric neutrino anomaly. 
For the degenerate case it may be allowed, 
 but may have a conflict with the constraint 
 $\sum_i m_{\nu i} \leq 1.8$ eV 
 due to observations of the large scale structures 
 in the present universe \cite{fukugita2}. 
The possibility $M_1 \gg 10^{10}$ GeV would be unlikely 
 in the view point of the re-heating temperature 
 of the inflationary universe. 
The condition for the inverse-decay can be consistent 
 with the neutrino oscillation data, and can be effective. 
In this case, there is the useful approximation formula \cite{K-T}: 
 $d \sim 1/(k \;  (\mbox{log} k)^{0.6})$ for $k > 1$, 
 where $k \sim  \Gamma_D/H(M_1)$. 
Considering the relation $d \sim 1/k \propto 1/m_{\nu 1}$, 
 the neutrino oscillation data and the cosmological constraint, 
 we can expect $d \geq10^{-3}$.  
As a result, both of the washing-out processes are 
 not so important for the resultant baryon asymmetry 
 in the conventional leptogenesis scenario. 

Now let us study how the above conclusion 
 is changed in models with the multi-Higgs doublets. 
For simplicity, we consider the model 
 with two up-type Higgs doublets, $H_1^u$ and $H_2^u$. 
This simple model includes all the essential points of our discussion. 
The Dirac neutrino mass matrix is written 
 by the sum of two Higgs doublets $H_i^u$,
\begin{eqnarray}
 m_D = h_1 \langle H_1^u \rangle + h_2 \langle H_2^u \rangle 
 = (h_1 \; \mbox{cos} \beta + h_2 \; \mbox{sin} \beta )
 \; v \; ,
\end{eqnarray}
where $v$ and $\beta$ are defined as 
 $v^2 = \langle  H_1^u \rangle^2 + \langle H_2^u \rangle ^2$ and 
 $\mbox{tan} \beta = \langle H_2^u \rangle /\langle H_1^u \rangle $, 
 respectively.
In the following analysis, we take $v=174$ GeV 
 as in the conventional scenario, for definiteness. 
It is straightforward to extend all the formulas necessary 
 in the leptogenesis scenario to the ones in the multi-Higgs case. 
For example, the CP-violating parameter Eq.~(\ref{epsilon}) 
 is modified to be the sum of 
\begin{eqnarray}
 \epsilon_a = \frac{1}{16\pi (h_a^\dagger h_a)_{11}} 
 \sum_b\sum_{j}\mbox{Im} \left[
 (h_a^\dagger h_b)_{1j} (h_b^\dagger h_a)_{1j} 
 +2 (h_a^\dagger h_a)_{1j} (h_b^\dagger h_b)_{1j} \right] \; 
  \frac{M_j}{M_1} \; 
\end{eqnarray}
for $ M_1 \ll M_j$, 
 where the first (second) term corresponds to the vertex 
 (wave function) correction. 
The crucial modification in the multi-Higgs models 
 is that the direct relation 
 between the Yukawa couplings and the Dirac mass matrix 
 to be compared with the experimental data is no longer valid, 
 because of the new parameter $\beta$ introduced. 
Even if the mass matrices, 
 $h_1 \langle H_1^u \rangle$ and $h_2 \langle H_2^u \rangle$, 
 are fixed by some predictive model and/or experiments,  
 there is still freedom to change the Yukawa couplings   
 according to $\beta$. 
Since general discussion about the multi-Higgs case 
 is very complicated, 
 we assume that $h_1 \langle H_1^u \rangle  \ll 
 h_2 \langle H_2^u \rangle$ and $h_1 \ll h_2$ 
 to make the essential point of our discussion clear. 
These assumption reduces our formulas 
 to the one in the conventional scenario 
 with identification $h = h_2$. 
However, note that the quantity defined as $\overline{m}_D = h v$ 
 is not the Dirac mass to be compared with the experiments, 
 and no longer suffer from the neutrino oscillation data. 
To avoid confusion, we use the notation, 
 $ (h^\dagger h)_{ii} v^2 /M_i = 
 (\overline{m}_D^\dagger \overline{m}_D)_{ii}/M_i 
 \sim \overline{m}_{\nu i}$, in the following. 
Note again that $\overline{m}_{\nu _i}$ 
 is not the physical mass eigenvalue. 

In the conventional scenario, the direct correspondence 
 between the Yukawa coupling and the Dirac mass matrix 
 allows us to describe the CP-violating parameter as 
\begin{eqnarray}
 \epsilon = \frac{3}{16 \pi (m_D^\dagger m_D)_{11}} \; 
 \sum_j \mbox{Im} \left[ 
 (m_D^\dagger m_D)_{1j}^2  \right] 
 \; \frac{M_j}{M_1}  \; .
\end{eqnarray}
by using the Dirac mass matrix. 
Suppose that $m_D$ is fixed by some mass matrix models and/or experiments, 
 we can evaluate $\epsilon$. 
The resultant $\epsilon$ and $Y_B \sim \epsilon/g_{*}$  
 may reveal a large discrepancy against the observed baryon asymmetry.  
However, note that, in our case, the above formula is modified as
 $\epsilon \rightarrow \epsilon/ \mbox{sin}^2 \beta$ through 
 $m_D \rightarrow \overline{m}_D = m_D/\mbox{sin} \beta$. 
Therefore, in the case with too small $\epsilon$, 
 we can easily reproduce the observed baryon asymmetry 
 by adjusting small $\sin\beta$ appropriately. 
This is the straightforward consequence according to 
 the introduction of new free parameter $\beta$. 
On the other hand, if $\epsilon$ is too large, 
 there seems to be no hope to obtain 
 the baryon asymmetry consistent with the observation.  
However, we will show that, in this case, 
 the washing-out processes can play an important role 
 so as to reduce the resultant $Y_B$ to the observed values. 
Since we no longer suffer from the neutrino oscillation data, 
 the Yukawa coupling $h=h_2$ can be taken to be large enough 
 for the washing-out process to become very effective. 
This is a new scenario in the leptogenesis 
 which has neither been considered seriously 
 nor be able to be realized in the conventional scenario 
 because of the constraints from the neutrino oscillation data. 

Now let us discuss the washing-out processes in detail 
 by integrating the Boltzmann equations out. 
In the conventional scenario, 
 numerical analysis have been done in detail \cite{luty} \cite{plumacher}, 
and all the quantities we need in this analysis are collected there. 
Thus, we do not discuss about derivations of formulas etc. in this letter. 
All our formulas are based on Ref.~\cite{plumacher}. 
We are interested in the parameter region 
 where the washing-out processes are effective. 
Although the conventional analysis has been performed 
 also on this parameter region, 
 it has not been seriously taken into account,   
 because of the incompatibility 
 with the neutrino oscillation data as discussed above. 

The Boltzmann equations are written by \cite{plumacher} 
\begin{eqnarray}
\frac{dY_{N_1}}{dz} &=& 
- \frac{z}{s H(M_1)} 
\left( \frac{Y_{N_1}}{Y_{N_1}^{eq}}-1 \right) \;
\gamma_D \; , \label{boltz1} \\
\frac{dY_{B-L}}{dz} &=& 
-\frac{z}{s H(M_1)} \left[  \left\{ 
\epsilon \left( \frac{Y_{N_1}}{Y_{N_1}^{eq}}-1 \right) 
+ \frac{1}{2} \frac{Y_{B-L}}{Y_l^{eq}} \right\} \gamma_{D_1} 
+ 2 \frac{Y_{B-L}}{Y_l^{eq}} 
 \left( \gamma_N + \gamma_{Nt} \right) \right] \; , 
\label{boltz2}
\end{eqnarray}
where $z=M_1/T$, and $s$ is the entropy density. 
Here $\gamma_D$ is given by 
\begin{eqnarray}
 \gamma_D \equiv n_{N_1}^{eq}\frac{K_1(z)}{K_2(z)} \Gamma_D
\end{eqnarray}
with the usual decay width of $N_1$ in the rest system, 
\begin{eqnarray}
\Gamma_D=\frac{(h^\dagger h)_{11}}{8\pi} M_1 
\sim \frac{\overline{m}_{\nu 1}}{8 \pi} \; \frac{M_1^2} {v^2} \; ,
\end{eqnarray}
the modified Bessel functions $K_i$, 
 and the number density $n^{eq}_{N_1} = M_1^3/(\pi^2 z) K_2(z)$. 
Here $\gamma_N$ ($\gamma_{Nt}$) corresponds to 
 the lepton-number-violating scattering process 
 $l \phi \leftrightarrow l \phi $ ($ l l \leftrightarrow \phi  \phi $) 
 through the effective dimension five interaction  of Eq.~(\ref{4point}). 
We have omitted the Higgs mediated processes for simplicity. 
These processes have an effect similar to that of the inverse-decay, 
 since both are classified in the same washing-out process 
 with the $N_1$ external leg and have amplitudes proportional 
 to $(h^\dagger h)_{11}$. 
For two body scattering, $\gamma$ is given by 
\begin{eqnarray} 
\gamma (a,b \leftrightarrow i,j)
= \frac{T}{64 \pi^4} 
 \int_{(m_a+m_b)^2}^\infty ds \hat{\sigma}(s) \sqrt{s} 
 K_1 \left( \frac{\sqrt{s}}{T} \right)  \; , 
 \label{twobody}
\end{eqnarray}
with the reduced cross section $\hat{\sigma}(s)$. 
Assuming, again, that the dominant contribution 
 is due to the single $N_i$ exchange for fixed $i$, 
 the sum of the reduced cross section for the scattering 
 $ l \phi \leftrightarrow l \phi$ and $ l l \leftrightarrow \phi \phi$ 
 is given by 
\begin{eqnarray}
\hat{\sigma}_{N+Nt} (s) 
= \hat{\sigma}_{N} (s) + \hat{\sigma}_{Nt} (s) 
\sim  \frac{3}{2 \pi}  \sum_j 
\frac{(\overline{m}_D^\dagger \overline{m}_D)_{jj}^2}{ v^4} \; 
 \frac{s}{M_j^2}  
\sim   \frac{3}{2 \pi} \frac{\overline{m}_{\nu 3}^2}{v^4} s \; . 
\end{eqnarray}
Here, we have further assumed the hierarchy 
 $\overline{m}_{\nu 1,2} \ll \overline{m}_{\nu 3}$ for simplicity. 
This also means that the washing-out process mediated by $N_3$ is dominant. 
Substituting the reduced cross section into Eq.~(\ref{twobody}), we obtain 
\begin{eqnarray}
 \gamma_{N+Nt} = \gamma_N + \gamma_{N t} 
 \sim \frac{3}{4 \pi} \; 
 \frac{\overline{m}_{\nu 3}^2 M_1^6}{v^4} \; z^{-6} \; .
\end{eqnarray}
This result is a good approximation 
 for the region $M_Z \ll T \leq M_1 \ll  M_3$ 
 which we are interested in, 
 where $M_Z$ is the Z-boson mass. 

Now we are ready to numerically analyze the Boltzmann equations. 
Given $\epsilon$, $\overline{m}_{\nu 1}$ and $\overline{m}_{\nu 3}$, 
 the Boltzmann equations can be easily integrated out. 
For very small $\overline{m}_{\nu 1}$ and $\overline{m}_{\nu 3}$, 
 we can find $Y_{B-L} \sim \epsilon/g_*$. 
Our interest is on the parameter region 
 where the washing-out processes play the important roles. 
Now we are free from the neutrino oscillation data, 
 and can take large input values 
 for $\overline{m}_{\nu 1}$ and $\overline{m}_{\nu 3}$. 
Let us consider two spacial cases. 
One is that the washing-out process through the inverse-decay is dominant.  
This case has been taken into account in the previous works 
 with the range of the resultant dilution factor $d \geq 10^{-3}$. 
The other is that the resultant $Y_{B-L}$ is dramatically reduced 
 by the washing-out through the dimension five interactions. 
For each cases, $Y_{B-L}$ is depicted in Fig.~1 and Fig.~2, respectively, 
 as a function of $\mbox{Log}_{10} (z)$.%
\footnote{ 
The region of the Yukawa couplings in the first case is limited. 
For large Yukawa couplings, 
the washing-out process through the dimension five interactions  
usually becomes important, and dominates 
(see Eqs.(\ref{appD}) and (\ref{appN}), 
and compare them). 
}
Here, we have taken the input parameter $\epsilon=10^{-3}$ as an example. 
The resultant dilution factor $d$ is independent 
 of this input in this numerical analysis. 
We can see that there is the dramatic reduction  
 of the resultant baryon number in each graph. 
 
There are useful approximation formulas  for two cases discussed above. 
The Boltzmann equation of Eq.~(\ref{boltz2}) can be rewritten as 
\begin{eqnarray}
\frac{dY_{B-L}}{dz} + P(z)Y_{B-L} = Q(z) \; ,  
\end{eqnarray}
where 
\begin{eqnarray}
P(z) = \frac{z}{s H(M_1) Y_l^{eq}} 
\left(   \frac{1}{2} \gamma_D + 2 \gamma_{N+Nt} \right) 
\equiv P_D(z) + P_N(z) 
\end{eqnarray}
with 
\begin{eqnarray}
 P_D(z) &=& 
 \frac{\overline{m}_{\nu 1} M_1^2}{32 \pi H(M_1) v^2}z^3 K_1(z) \; , \\
 P_N(z) &=& 
 \frac{3 \overline{m}_{\nu 3}^2 M_1^3}{4 \pi^3 H(M_1)v^4} z^{-2}  \; , 
\label{PN}
\end{eqnarray}
and 
\begin{eqnarray} 
Q(z) = - \epsilon \; \frac{z}{s H(M_1)} 
\left( \frac{Y_{N_1}}{Y_{N_1}^{eq}}-1 \right)  \gamma_D  \; . 
 \end{eqnarray} 
Here we have used $ s Y_l^{eq} = \frac{2}{\pi^2} M^3 z^{-3}$. 
The solution of the above Boltzmann equation 
 with the initial condition $Y_{B-L}(0) = 0 $ is given by 
\begin{eqnarray} 
 Y_{B-L}(z)=\int_{0}^z dx Q(x) e^{-\int_x^z P(y) dy } \;  
\label{sol}
\end{eqnarray} 
with the solution of Eq.~(\ref{boltz1}). 
We can find that $Y_{B-L}(\infty) \sim \epsilon/g_* $ if $P(z) \ll 1 $. 

Noting that the out-of-equilibrium decay begins roughly at $T \sim M_1$, 
 the condition $P(1) \geq 1$ gives a reasonable condition 
 for the washing-out processes to be effective. 
For the above two cases, we can obtain 
 the conditions on the light neutrino mass eigenvalues such that
\begin{eqnarray}
 P_D(1) \geq 1 ~~& \rightarrow &~~  
 \overline{m}_{\nu 1}\geq  4 \times 10^{-3} \mbox{eV}  \; , 
 \label{PDbound} \\ 
 P_N(1) \geq 1 ~~& \rightarrow &~~ 
 \overline{m}_{\nu 3} \geq  2~\mbox{eV}~ 
 \left( \frac{10^{10}\mbox{GeV}}{M_1} \right)^{1/2} \; .  
\label{PNbound}
\end{eqnarray}
These results are in good agreement with the numerical calculations 
 as can be seen in Fig~1 and 2. 

The washing-out processes play 
 the important roles in the case $P(1) \gg 1$. 
Using the method of the steepest decent in Eq.~(\ref{sol}), 
we can find \cite{K-T}  
\begin{eqnarray}
 Y_{B-L}(\infty) \sim \frac{\epsilon}{g_*} \; 
 a^{-1/2} z_f^{3/2} 
 \mbox{exp} \left[
 -z_f -\int_{z_f}^{\infty} dz P(z) \right] \; ,  
\end{eqnarray}
where $a = - d P(z)/dz |_{z=z_f} (>0)$ with $z_f$ defined as $P(z_f) =1$.   
In the first case, $P_D(1) \gg 1$ but $P_N (1) \ll 1$, 
 it is difficult to obtain a simple formula 
 because of the Bessel function in $P_D(z)$. 
Although we can find a rough formula $ Y_{B-L} \propto 1/k$, 
 more reasonable fitting formula is known such as \cite{K-T} 
\begin{eqnarray}
 Y_{B-L}(\infty) \sim \frac{\epsilon}{g_*} \times 
 \frac{0.1}{k (\mbox{ln} k)^{0.6}} 
\label{appD}
\end{eqnarray}
with $k \sim P_D(1)$. 
In the other case, $P_D(1) \ll 1$ but $P_N (1) \gg 1$, 
 analytic calculation is possible by using Eq.~(\ref{PN}), 
 and the approximation formula is found to be 
\begin{eqnarray}
 Y_{B-L}(\infty) \sim \frac{\epsilon}{g_*}  
  \times \Delta \mbox{exp}\left( -2 \sqrt{\Delta} \right)  
\label{appN}
\end{eqnarray}
with 
\begin{eqnarray}   
 \Delta = 
\frac{3 \overline{m}_{\nu 3}^2 M_1^3} {4 \pi^3 H(M_1)v^4} \; .
\end{eqnarray}   
We can check that the both formulas 
 give good approximations compared to the numerical results. 

We have learned that there is the parameter region 
 where the washing-out processes become important. 
Here, remember that $\epsilon$ is the function of the Yukawa couplings 
 and becomes large according to the power law, $\epsilon \propto h^2$. 
In the case $P_D(1) \gg 1$ but $P_N(1) \ll 1$, 
 from Eq.~(\ref{appD}), we can expect that 
 the resultant baryon number is saturated to a constant, 
 as Yukawa coupling becomes large.  
On the other hand, 
 in the case $P_N(1) \gg 1$ but $P_D(1) \ll 1$, 
 Eq.~(\ref{appN}) leads to the conclusion 
 that the resultant baryon number becomes 
 an exponentially decreasing function 
 of the (large) Yukawa couplings. 
This implies that, even if the parameter $\epsilon/g_*$ itself 
 is too large to be consistent with the observation, 
 new parameter region of large Yukawa couplings comes out, 
 so that the observed baryon number can be reproduced. 
This is a new interesting scenario. 

In summary, we have studied the leptogenesis scenario 
 in models with multi-Higgs doublets. 
In our scenario, the direct correspondence 
 between the Yukawa coupling and the mass matrix is lost.
As a result, new parameter region can be consistent 
 with the neutrino oscillation data. 
We have pointed out the importance of the washing-out process  
 which has not been seriously taken into account 
 in the conventional scenario. 
We found a new parameter region of the large Yukawa couplings  
 which can reproduce the observed baryon number 
 through the exponential dumping of 
 too large CP-violating parameter $\epsilon$ 
 due to the washing-out processes 
 through the dimension five interactions . 

Finally, we give a comment on the application of our scenario 
 to some neutrino mass matrix models. 
Many models of the Majorana neutrinos with the see-saw mechanism 
 have been proposed, which predict and/or reproduce the observed 
 neutrino properties. 
When the leptogenesis scenario is taken into account in such models, 
 some of them may predict too large baryon asymmetry 
 to be consistent with the observations. 
This is very problematic in cosmology.%
\footnote{
If we assume that the re-heating temperature of 
 the inflationary universe is lower 
 than the right-handed neutrino masses, 
 this problem can be easily avoided. 
However, the leptogenesis no longer works. }
However, if such a model is the multi-Higgs doublet model, 
 the scenario we have discussed in this letter 
 is applicable and can make the model viable. 
Further discussion needs a concrete model, 
 and such a model is worth investigating. 

The authors would like to thank 
 Rabindra Mohapatra and Markus Luty 
 for useful discussions. 

\newpage

\newpage
\begin{figure}
\begin{center}
\epsfig{file=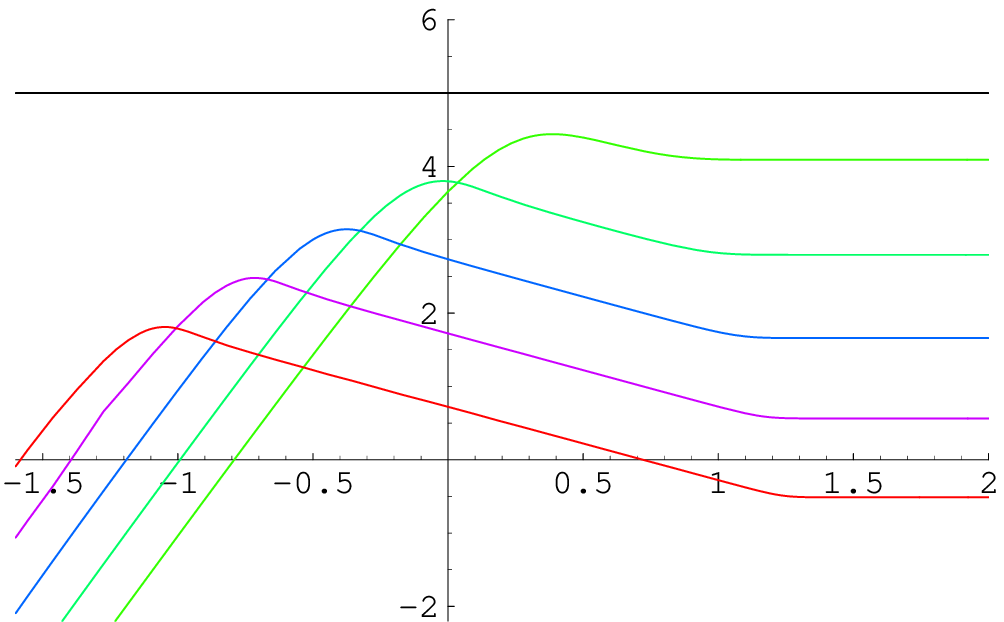, width=12cm}
\caption{
The solution of the Boltzmann equation 
($\mbox{Log}_{10}(Y_{B-L} \times 10^{10})$) 
with $\epsilon/g_*=10^{-5}$ 
(the upper horizontal line), 
$M_1 =10^{10}$ GeV and $\overline{m}_{\nu 3}=0.2 $ eV. 
The solutions for 
$\overline{m}_{\nu 1}=4 \times 10^{-3}$, 
$0.04$, $0.4$, $4$ and $40$ eV 
are plotted from above at $z=10^2$. 
}
\end{center}
\end{figure}

\begin{figure}
\begin{center}
\epsfig{file=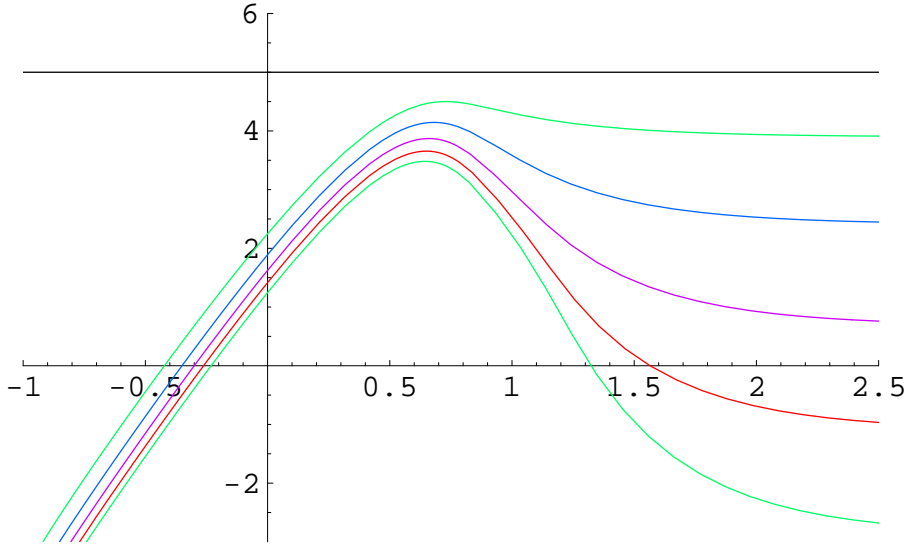, width=12cm}
\caption{
The solution of the Boltzmann equation 
($\mbox{Log}_{10}(Y_{B-L} \times 10^{10})$) 
with $\epsilon/g_*=10^{-5}$, $M_1 =10^{10}$ GeV 
and $\overline{m}_{\nu 1}=4 \times 10^{-4}$ eV.  
The solutions for the input values 
$\overline{m}_{\nu 3}= 7$, $12$, $17$, $22$ and $27$ eV 
are plotted from above. 
}
\end{center}
\end{figure}

\end{document}